\documentclass[aps,pre,twocolumn,groupedaddress,showpacs,floatfix,amsmath]{revtex4}
\usepackage{graphicx,amsmath}
\begin{document}
\preprint{}
\title{ 
Qualitative difference in rheology between fragile and network-forming strong liquids}
\author{
Akira Furukawa}
\email{furu@iis.u-tokyo.ac.jp}
\affiliation{Institute of Industrial Science, University of Tokyo, Meguro-ku, 
Tokyo 153-8505, Japan. }
\date{\today}
\begin{abstract}
 We elucidate a qualitative difference in rheology between fragile and network-forming strong liquids. In a flow field, the structural configuration is distorted in accordance with the flow symmetry, whereas the form of the interaction potential remains unchanged. The role of this mismatch in the relaxation mechanism under the flow field is crucial for understanding the shear-thinning mechanism and differs between strong and fragile glass formers. In fragile glass formers, shear thinning can be attributed to the shear-induced reduction of the {\it effective density}. In contrast, in strong glass formers, the shear-induced reduction of the {\it effective activation energy} is a possible origin of  a significant acceleration in relaxation. Our simple predictions of the crossover shear rate, $\dot\gamma_{\rm c}$, from Newtonian to non-Newtonian behaviors can be expressed in terms of experimental observables: in fragile liquids, $\dot\gamma_{\rm c}=(\rho \partial \tau_\alpha/\partial \rho)^{-1}$, where $\rho$ and $\tau_\alpha$ are the density and structural relaxation time, respectively, and in strong liquids, $\dot\gamma_{\rm c}= (\tau_\alpha\Delta E_0 /T)^{-1} $, where $T$ and $\Delta E_0$ are the temperature and equilibrium activation energy, respectively. These predictions are consistent with the results of molecular dynamics simulations for four different glass formers: two fragile and two strong ones. This different route to the non-Newtonian flow response is related to differences in the role of density in the relaxation dynamics. 
\end{abstract}

\maketitle

\section{Introduction}
 Shear thinning is one of the most ubiquitous non-Newtonian flow behaviors in glassy materials \cite{Yamamoto-Onuki,Berthier-Barrat,Varnik,Shi-Falk,Lemaitre,FurukawaS1,Webb-Dingwell,Kato-Kawamura-Inoue-Chen,Lu-Ravichandran-Johonson,Besseling-Isa-Ballesta-Petekidis-Cates-Poon,Liu-NagelB,LarsonB,dynamic_heterogeneityB,VoightmannR}: when an imposed shear rate  $\dot\gamma$ is smaller than the crossover value $\dot\gamma_{\rm c}$, the shear viscosity $\hat\eta$ and structural relaxation time $\hat\tau_\alpha$ under the flow remain the same as those in equilibrium ($\dot\gamma=0$), $\eta$ and $\tau_\alpha$, respectively. In contrast, when $\dot\gamma > \dot\gamma_{\rm c}$, $\hat\eta$ and $\hat\tau_\alpha$  decrease significantly with increasing $\dot\gamma$. This non-linear flow response usually develops into more complex phenomena, such as shear-banding and fracture, drastically altering the mechanical properties. Thus, understanding and controlling the shear thinning behavior are of particular importance in the design of the processing of glassy materials. However, although a wide variety of models have been proposed to describe the shear thinning of glass-forming liquids (see papers \cite{Spaepen,Argon,Falk-Langer,Fuchs-Cates,Brader-Cates-Fuchs,Miyazaki-Reichman,SGR,Otsuki-Sasa,Lubchenko,Trond-Tanaka} and the references therein), there is still no general consensus regarding the underlying mechanism. 

In most experiments and simulations of non-polymeric glassy liquids, the following characteristic rheological features appear around the crossover from Newtonian to non-Newtonian flow behaviors ($\dot\gamma\sim \dot\gamma_{\rm c}$):

{\bf (1).-   Very large time-scale separation $\dot\gamma_{\rm c}\tau_\alpha\ll 1$}: Shear thinning starts when $\dot\gamma$ is several orders of magnitude smaller than $1/\tau_\alpha$ \cite{Webb-Dingwell,Kato-Kawamura-Inoue-Chen,Lu-Ravichandran-Johonson,Yamamoto-Onuki,Berthier-Barrat,Varnik,FurukawaS1,Lemaitre}, which may exclude the possibility of the usual constitutive instability \cite{FurukawaS1,Lubchenko}. 

{\bf (2).-   Very small composition inhomogeneity}: An enhancement of density fluctuations is highly suppressed \cite{Yamamoto-Onuki,Lu-Ravichandran-Johonson}, indicating that shear thinning is not triggered by a shear-induced phase transition or composition inhomogeneity, as has frequently been observed in soft matter systems  \cite{Onuki,Helfand-Fredrickson,Milner,Imaeda-Furukawa-Onuki,FurukawaS2}.

To provide a logical explanation of these characteristic features, we recently  proposed a different theoretical model \cite{FurukawaS3}. Details of the basic model components are explained here: 

{\bf (i).-  Flow-induced deformation}: During the structural relaxation period, the particle configurations, on average, are preserved. Thus, in a flow field with shear rate $\dot\gamma$ the average structure undergoes shear deformation with the strain of $\gamma=\dot\gamma\hat\tau_\alpha$. 

{\bf (ii).-  Reduction of the effective density}: 
The average configuration is distorted in accordance with the given flow symmetry.  In the flow, to sustain the average stress, the neighboring particles show greater overlap along the compression axis than at equilibrium. In other words, the effective particle ``core'' size is reduced along this direction. In contrast, along the elongation axis, the particles are diluted, which does not indicate an increase of the particle ``core'' size. Due to this asymmetric flow effect, the density is {\it effectively} reduced.

{\bf (iii).- Acceleration of the relaxation}: 
Near the glass transition point, because the density dependence of the structural relaxation is quite severe, even a very small decrease in the effective density can  decrease the relaxation time significantly; the shear-induced small reduction of the effective density is responsible for the marked shear-thinning. 

These situations, (i)-(iii), are schematically illustrated in Fig. 1.

\begin{figure}[th] 
\includegraphics[width=.48\textwidth]{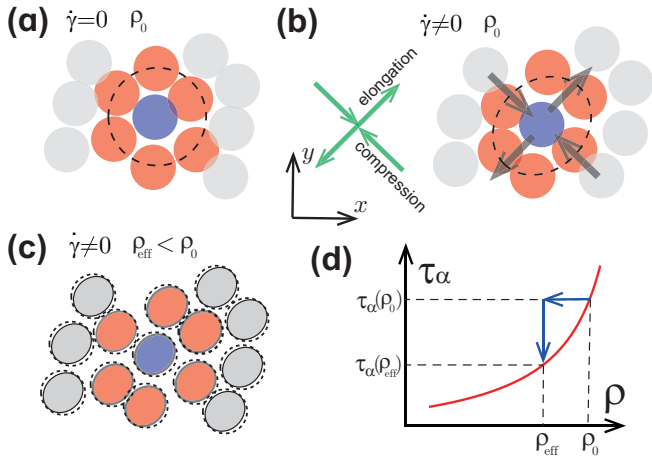}
\caption{(Color online) A schematic of the shear thinning mechanism for fragile glass formers proposed in Ref. \cite{FurukawaS3}. (a) Particle configuration at equilibrium ($\dot\gamma=0$). In a high-density liquid, the neighboring particles overlap significantly due to thermal fluctuations; the distance separations among them strongly fluctuate in space and time. (b) Under the shear flow, Eq. (\ref{shear-flow}), the average configuration is elongated and compressed along $x=y$ and $x=-y$, respectively. (c) Along $x=-y$, to sustain the average shear stress, the “core” size, below which the particles can hardly be closer to each other, is smaller than at equilibrium. On the other hand, along $x = y$, the ``core'' size remains unchanged. 
The dashed and solid lines represent the effective core size at $\dot\gamma=0$ and $\dot\gamma\ne 0$, respectively.  
This asymmetric flow effect (exaggerated in the illustration) leads to a decreased effective density $\rho_{\rm eff}$. 
(d) For $\gamma\ll 1$, the dynamics can be assumed to be mapped onto the equilibrium dynamics. For fragile glass formers, because $\tau_\alpha$ depends strongly on the density, even a very small reduction in $\rho_{\rm eff}$ significantly accelerates the relaxation.  }
\label{Fig1}
\end{figure}

 In Ref. \cite{FurukawaS3}, it was shown that a model based on the above perspective agrees with the simulation results for a two-dimensional model glass former. However, in the above argument, particularly for (ii) and (iii) (see the comment in \cite{comment1}), the considered system is implicitly assumed to be a fragile glass-former. Additionally, the model glass former used for the simulations does show the typical properties of fragile liquids. Glass-forming liquids are categorized into two classes: ``strong'' and ``fragile'' \cite{Angell}. The density and temperature dependencies of the structural relaxation time $\tau_\alpha$ of strong glass formers are quite different from those of fragile glass formers, suggesting a difference in the relaxation mechanism between these classes of glass formers \cite{Vogel-Glotzer,Coslovich-Pastore,Kim-Saito,Staley-Flenner-Szamel,FurukawaG1}. Thus, although both strong and fragile glass formers show very similar $\dot\gamma$-$\hat\eta$ curves, this does not mean that the two classes of glass formers share a common underlying shear-thinning mechanism. 

\section{Difference in shear distortions}

In this study, we investigate an essential difference in the rheological responses between fragile and strong glass formers by investigating the shear distortions of the average particle configurations. Furthermore, we discuss a possible difference in the shear-thinning mechanism between these two classes of glass formers. To this end, we perform molecular dynamics simulations using four popular three-dimensional models of glass-forming binary mixtures: The model strong glass formers are the van Beest-Kramer-van Santen (BKS) \cite{van Beest-Kramer-van Santen} and the Coslovich-Pastore (CP) \cite{Coslovich-Pastore} models. Both models are often used as models for amorphous and supercooled silica (Si{$\rm O_2$}), which is the prototypical strong glass former. The fragile glass formers are the Kob-Andersen (KA) \cite{Kob-Andersen} and the Bernu-Hiwatari-Hansen (BHH) soft-sphere \cite{Bernu-Hiwatari-Hansen} models. The latter is more fragile than the former. Details of the simulations and models are presented in the Appendix. In the following investigation, we consider the simple shear flow without the loss of generality: 
\begin{eqnarray}
\langle {\mbox{\boldmath$v$}}\rangle={\dot \gamma}y\hat{\mbox{\boldmath$x$}}, \label{shear-flow}
\end{eqnarray}
where the $x$ axis is along the direction of the mean flow, the $y$ axis is along the mean velocity gradient, and $\hat{\mbox{\boldmath$x$}}$ ($\hat{\mbox{\boldmath$y$}}$) is the unit vector along the $x$ ($y$) axis. Here, ${\mbox{\boldmath$v$}}$ is the velocity field, and $\langle \cdots\rangle$ denotes the spatial average.

\begin{figure}[bt] 
\includegraphics[width=.47\textwidth]{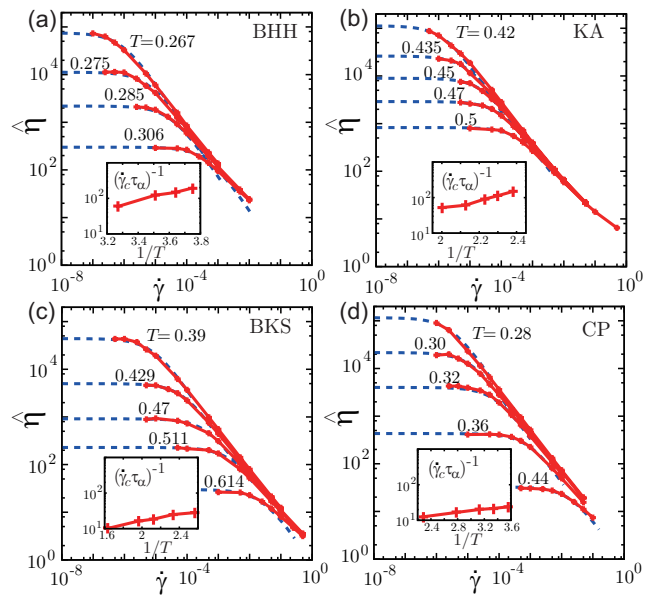}
\caption{ 
(Color online) The $\dot\gamma$-dependent steady-state shear viscosity $\hat\eta(\dot\gamma)$ for the BHH (a), KA (b), BKS (c), and CP (d) models. The crossover shear rate $\dot\gamma_{\rm c}$ is determined by the fit, $\eta/(1+\dot\gamma/\dot\gamma_{\rm c})$, represented by the blue dashed lines. Here, $\eta$ is the viscosity at equilibrium. In the insets, we plot the inverse of the degree of the shear ``distortion'' at the crossover, $(\dot\gamma_{\rm c}\tau_\alpha)^{-1}$, against $1/T$.}
\label{Fig2}
\end{figure}

The $\dot\gamma$-$\hat\eta$ curves for the four model glass formers are shown in Fig. 2, and all curves exhibit very similar shear-thinning behavior. However, focusing on the average degree of the shear ``distortion'' at the crossover, $\dot\gamma_{\rm c}\tau_\alpha$, reveals a significant difference.  In the insets of Figs. 2(a)-(d), we plot $(\dot\gamma_{\rm c}\tau_\alpha)^{-1}$ against the inverse of temperature, $1/T$, where $\tau_\alpha$ is evaluated by the shear-stress autocorrelation \cite{comment_relaxation}. 
Over the range of the degree of supercooling studied here, the value of $\dot\gamma_{\rm c}\tau_\alpha$ for fragile liquids is significantly smaller than that for strong liquids, indicating that for fragile liquids, the relaxation dynamics can be more easily accelerated at smaller distortions.  As we discuss below, this difference in $\dot\gamma_{\rm c}\tau_\alpha$ is a manifestation of the essential difference in the shear-thinning mechanism between strong and fragile glass formers.

To illustrate how the particle configuration is distorted by the shear flow, let us first investigate the azimuthal angular dependence of the nearest neighbor distance between the $\mu$- and $\nu$-species, which is defined as  
\begin{eqnarray}
\lambda_{\mu\nu}(\theta) =\dfrac{\int_0^{r_{\mu\nu}^{(0)}} dr \int_0^\pi d\psi \sin\psi r^3 g_{\mu\nu}(r,\psi,\theta)}{{\int_0^{r_{\mu\nu}^{(0)}} dr \int_0^\pi d\psi \sin\psi  r^2g_{\mu\nu}(r,\psi,\theta)}},
\end{eqnarray} 
where $g_{\mu\nu}(r,\psi,\theta)$ is the pair correlation function and $r_{\mu\nu}^{(0)}$ is the distance at which $g_{\mu\nu}$ reaches its first minimum. Here, $\psi=\cos^{-1}(z/r)$ and $\theta=\tan^{-1}(y/x)$. 
In Fig. 3, for the fragile BHH and strong BKS models, we plot $\lambda_{\mu\nu}(\theta)$ for several $\dot\gamma$ at around the crossover ($\hat\tau_\alpha\cong \tau_\alpha$). 
For the fragile BHH model, $\lambda_{\mu\nu}(\theta)$ behaves as 
\begin{eqnarray}
\lambda_{\mu\nu}(\theta) \cong \lambda_{\mu\nu}^{(0)} [ 1 + c_\lambda \dot\gamma \hat\tau_\alpha \sin (2\theta)],
\end{eqnarray}
where $c_\lambda$ is a numerical constant and $\lambda_{\mu\nu}^{(0)}$ is the nearest neighbor distance at equilibrium. Thus, the nearest-neighbor distance is elongated and compressed at $\theta=\pi/4$ ($x=y$) and $3\pi/4$ ($x=-y$), respectively, which is consistent with the spatial symmetry of the simple shear flow. In contrast, for the strong BKS model, $\lambda_{\mu\nu}(\theta)$ behaves very differently. $\lambda_{\rm OO}(\theta)$ exhibits similar behavior, but $\lambda_{\rm Si Si}(\theta)$ and $\lambda_{\rm Si O}(\theta)$ remain near their equilibrium values. 

\begin{figure}[hbt] 
\includegraphics[width=.5\textwidth]{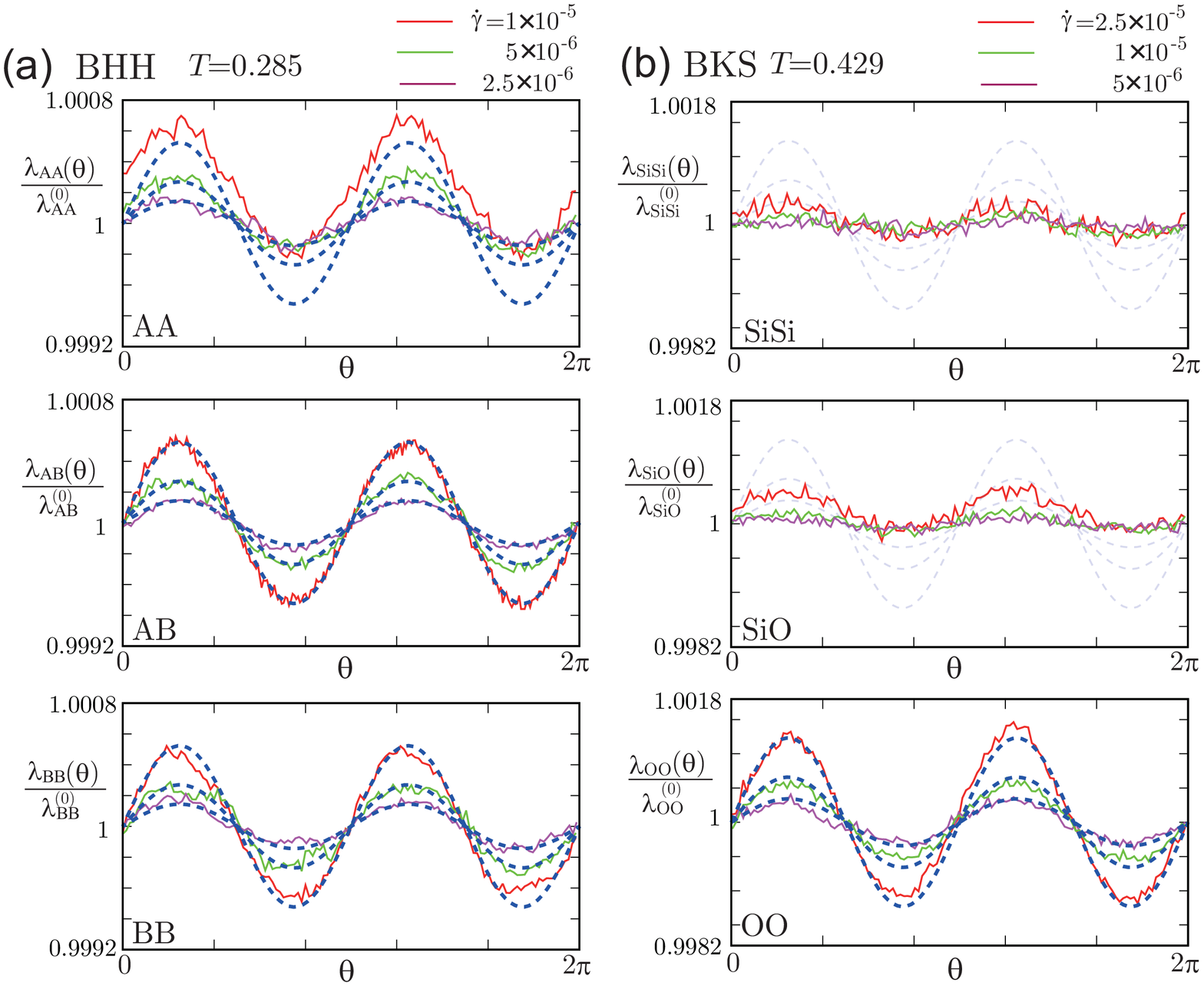}
\caption{ 
(Color online) $\lambda_{\mu\nu}(\theta)$ scaled by $\lambda_{\mu\nu}^{(0)}$ for several shear rates for the BHH (a) and BKS (b) models. 
Here, $\lambda_{\mu\nu}^{(0)}$ is the nearest neighbor distance at equilibrium. 
In (a), the blue dashed line represents $\lambda_{\mu\nu}^{(0)}[1+c_\lambda\dot\gamma \hat\tau_\alpha \sin(2\theta)]$ with $c_\lambda=0.25$. In (b), for the O-O pair, the blue dashed line represents $\lambda_{\rm OO}^{(0)}[1+c'_\lambda\dot\gamma\hat\tau_\alpha \sin(2\theta)]$ with $c'_\lambda=0.14$, while $\lambda_{\rm SiSi}$ and $\lambda_{\rm SiO}$ remain near their equilibrium values. }
\label{Fig3}
\end{figure}

This distinction in $\lambda_{\mu\nu}(\theta)$ can be understood by examining the ``distortion'' of the pair-correlation function. Before proceeding, we provide general remarks about the pair correlation under the simple shear flow. Due to the flow symmetry, the pair-correlation function is generally given as \cite{Kirkwood-Buff-Green}
\begin{eqnarray}
g_{\mu\nu}({\mbox{\boldmath$r$}})=g_{\mu\nu}^{(0)}(r)+g_{\mu\nu}^{(1)}(r) {\hat x}{\hat y} + \cdots, \label{g_of_r}
\end{eqnarray} 
where ${\hat x}=x/r=\sin\psi\cos\theta$ and ${\hat y}=y/r=\sin\psi\sin\theta$. 
Here, $g_{\mu\nu}^{(0)}(r)$ is the pair-correlation function at equilibrium ($\dot\gamma=0$) and $g_{\mu\nu}^{(1)}(r) {\hat x}{\hat y}$ is the leading order term of the deviation from $g_{\mu\nu}^{(0)}(r)$, which is responsible for the non-zero average shear stress, $\sigma_{xy}$, as  
\begin{eqnarray}
\sigma_{xy}&=& \dfrac{1}{2}\sum_{\mu,\nu} \rho_\mu\rho_\nu  \int d{\mbox{\boldmath$r$}} {\hat x}^2{\hat y}^2 r \dfrac{du_{\mu\nu}}{dr} g_{\mu\nu}^{(1)}(r), 
\label{shear_stress_R}
\end{eqnarray} 
where $\rho_\mu$ is the number density of $\mu$-specie particles and $u_{\mu\nu}$ is the interaction potential between $\mu$- and $\nu$-species.
Note that in silica, $\sigma_{xy}$ is dominated by the contributions from the Si-O pairs. 
In Fig. 4, for the fragile BHH and strong BKS models, $g_{\mu\nu}^{(1)}(r)$ near the crossover is shown. 
The behavior of $g_{\mu\nu}^{(1)}(r)$ in the fragile BHH model is well described by \cite{Hanley-Rainwater-Hess,Suzuki-Haimovich-Egami}
\begin{eqnarray}
g_{\mu\nu}^{(1)}(r)  \cong  - c_g \dot\gamma\hat\tau_\alpha  r\dfrac{\partial g_{\mu\nu}^{(0)}}{\partial r}, \label{configuration1}
\end{eqnarray}
where $c_g$ is a numerical constant of the order of unity. 
Eq. (\ref{configuration1}) can be understood as a consequence of balancing the mass-conserved advection and the relaxation in the steady state: 
\begin{eqnarray} 
\dot\gamma y \dfrac{\partial }{\partial x} g_{\mu\nu} \sim  -\dfrac{1}{\hat\tau_{\alpha}}(g_{\mu\nu}-g_{\mu\nu}^{(0)}).  \label{eq_motion1}
\end{eqnarray}
By taking the leading order of $\dot\gamma$, we obtain Eq. (\ref{configuration1}). 
As shown in Fig. 4(a), this approximate form of $g_{\mu\nu}^{(1)}$ with $c_g=0.65$ reproduces the simulation results. In contrast, in Fig. 4(c), we find that for the strong BKS model, the functional forms of $g_{\mu\nu}^{(1)}$ are quite different from those for the BHH model. In particular, the behavior of $g_{\rm Si O}^{(1)}$ is remarkably distinct. That is, there is no minimum in $g_{\rm Si O}^{(1)}$ at the nearest-neighbor distance. Instead, $g_{\rm Si O}^{(0)}$ and $g_{\rm Si O}^{(1)}$ share approximately the same position and width of their first peaks, as shown by the blue dashed line for $g_{\rm Si O}^{(1)}(r)$ in Fig. 4(c). Note that the same behaviors are found in the CP model. This behavior of $g_{\rm Si O}^{(1)}$ is described by the following empirical form: 
\begin{eqnarray}
g_{\rm Si O}^{(1)}(r) \cong h_g \dot\gamma \hat\tau_\alpha g_{\rm Si O}^{(0)}, \label{configuration2}
\end{eqnarray}
where $h_g$ is a constant of the order of unity. Eq. (\ref{configuration2}) shows that the relaxation term in the steady state is balanced by a different type of shear distortion: 
\begin{eqnarray} 
- \dot\gamma {\hat x}{\hat y}  g_{\rm Si O} \sim -\dfrac{1}{\hat\tau_\alpha}(g_{\rm Si O}-g_{\rm Si O}^{(0)}), \label{eq_motion2} 
\end{eqnarray}
where the l.h.s. of Eq. (\ref{eq_motion2}) is different from that in Eq. (\ref{eq_motion1}). 

\begin{figure}[bt] 
\includegraphics[width=.5\textwidth]{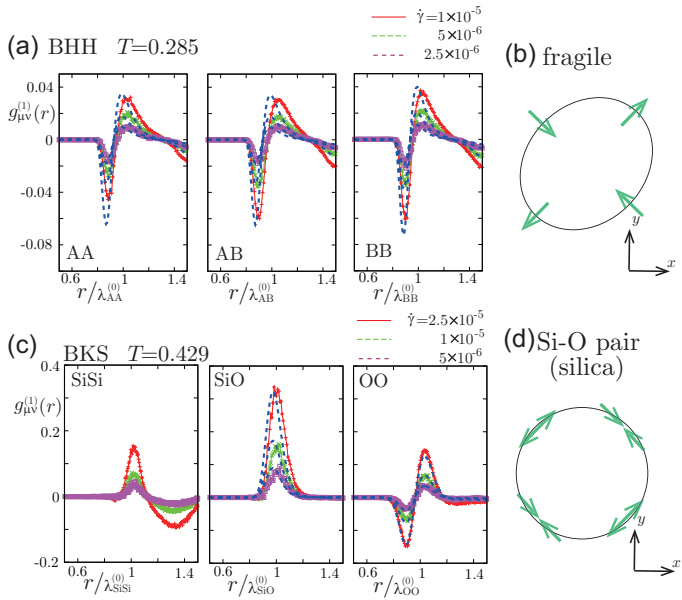}
\caption{ 
(Color online) $g_{\mu\nu}^{(1)}(r)$ for the BHH (a) and BKS (c) models in the Newtonian regime. In (a), the blue dashed line represents $-c_g \dot\gamma\hat\tau_\alpha r \partial g_{\mu \nu}^{(0)}/\partial r$ with $c_g=0.65$. In (c), for the Si-O pair, the blue dashed line represents $h_g \dot\gamma\hat\tau_\alpha g_{\rm SiO}^{(0)}$ with $h_g=3.5$, while for the O-O pair, the blue dashed line represents $-c_g' \dot\gamma\hat\tau_\alpha r \partial g_{\rm OO}^{(0)}/\partial r$ with $c_g'=0.65$.  
The schematic illustrations of the shear ``distortion'' show the following: Under the shear flow, the particle configurations in fragile liquids are elongated and compressed in $x=y$ and $x=-y$, respectively, as illustrated in (b).  In contrast, the average configuration of the Si-O pair ``flows'' with a preserved bond length, as indicated in (d).  }
\label{Fig4}
\end{figure}

We now discuss the physical origin of these differences in the steady-state particle configuration. Fragile glass formers are highly incompressible, and their density is locally conserved \cite{FurukawaG1}. In a sufficiently high-density fragile liquid, steric constraints due to the repulsive interactions at the nearest-neighbor distances dominate the blocking of the rearrangements for a long period, during which the average configurations are distorted in a conserved manner as $\nabla\cdot ({\mbox{\boldmath$v$}} g)=\dot\gamma y \partial g/\partial x$. This situation is schematically illustrated in Fig. 4(b). In contrast, for the strong silica, as demonstrated in Ref. \cite{FurukawaG1}, the density is not locally conserved (but is, of course, globally conserved), which may be due to the lower packing and the associated higher compressible nature than those of fragile liquids. Consequently, the density itself is less relevant for blocking particle rearrangements. Additionally, the covalent Si-O attraction is sufficiently strong. For these reasons, under the simple shear flow, the Si-O pair does not obey a mass-conserved advection, as schematically shown in Fig. 4(d). 
Note, however, that elongation and compression (by approximately $\gamma=\dot\gamma\hat\tau_\alpha$) in accordance with the flow symmetry occur in the SiO$_4$ tetrahedral units of a network structure, to which the distortions of the Si-Si and O-O pairs are subordinate: The Si-O and Si-Si pair correlations describing a ``frame'' structure  increase and decrease along $x=y$ and $x=-y$, respectively, preserving their nearest-neighbor distances. 
In contrast, the O-O pair determines the ``outer surfaces'' of the SiO$_4$ units, and its nearest neighbor distances are longer and shorter in $x=y$ and $x=-y$, respectively. 
These trends are indicated in $g_{\mu\nu}^{(1)}(r)$ and $\lambda_{\mu\nu}(\theta)$. A more detailed structural analysis will be presented elsewhere.

\section{Difference in shear-thinning mechanisms}

Based on the above investigations, we propose the shear-thinning mechanism and consider the difference between the mechanisms for fragile and strong liquids. 
 
\subsection{Fragile glass-formers: shear-induced reduction of the effective density}
We proposed a shear-thinning mechanism for fragile glass formers in Ref. \cite{FurukawaS3}. 
Here, we further verify its validity by simulating different model fragile glass formers. For a detailed argument, the reader is referred to Ref. \cite{FurukawaS3}. 
As demonstrated above, along $x=-y$, the neighboring particles show greater overlap than at equilibrium; specifically, the effective particle (core) radius is reduced by $\gamma=\dot\gamma\hat\tau_\alpha$. In contrast, along $x=y$, the particles are diluted by $\gamma$, but the effective radius itself does not expand. Consequently, the effective density is decreased by approximately $\gamma$ as 
\begin{eqnarray}
\rho_{\rm eff} (\dot\gamma) = \rho(1-a \gamma), \label{effective_density}
\end{eqnarray}
where $a$ is a constant of the order of unity and depends on the particle shape and $\rho$ is the equilibrium density. 
Note that such a reduction of $\rho_{\rm eff}$ cannot be directly reflected in the change of $g_{\mu\nu}({\mbox{\boldmath$r$}})$ because generally, the true density itself cannot be reduced by simple shear. It was shown in simulations \cite{Yamamoto-Onuki,Miyazaki-Yamamoto-Reichman}, that for $\gamma\ll 1$ the (deviatoric) dynamics \cite{comment2} are remain almost isotropic; thus, we can suppose that the relaxation time under the shear flow, $\hat\tau_\alpha$, is mapped onto the equilibrium one, $\tau_\alpha$, according to 
\begin{eqnarray}
\hat\tau_\alpha(\rho,T,\dot\gamma) = \tau_\alpha(\rho_{\rm eff},T),  \label{mapping_fragile}
\end{eqnarray}
where the flow effect is taken into account through the effective density. This equation is essentially nonlinear in $\dot\gamma$. 
For $\gamma\ll 1$, by expanding $\tau_\alpha(\rho_{\rm eff},T)=\tau_\alpha[\rho(1-a\gamma),T]$ in $\gamma$, we obtain 
\begin{eqnarray}
\hat\tau_\alpha(\rho,T,\dot\gamma) = \dfrac{\tau_\alpha(\rho,T)}{1+a \dot\gamma \rho \dfrac{\partial \tau_\alpha}{\partial \rho}}. \label{leading_fragile}
\end{eqnarray}
Therefore, $\dot\gamma_{\rm c}$ is given by 
\begin{eqnarray}
\dot\gamma_{\rm c} \cong \biggl( \rho \dfrac{\partial \tau_\alpha}{\partial  \rho}\biggr)^{-1}\label{crossover_fragile}. 
\end{eqnarray} 
Figs. 5(a) and (b) show that Eq. (\ref{crossover_fragile}) describes the crossover behaviors of the BHH and KA models well.  

\begin{figure}[tb] 
\includegraphics[width=.45\textwidth]{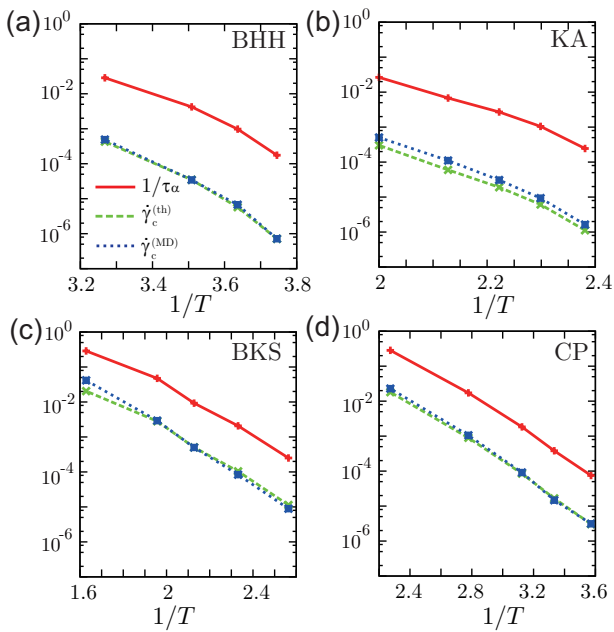}
\caption{ 
(Color online) The crossover shear rate $\dot\gamma_{\rm c}^{\rm (th)}$ determined by the theoretical model and $\dot\gamma_{\rm c}^{\rm (MD)}$ estimated from the simulation results are plotted against $1/T$ for the BHH (a), KA (b), BKS (c), and CP (d) models. The inverse of $\tau_\alpha$ is also represented. }
\label{Fig}
\end{figure}

\subsection{Strong silica: shear-induced reduction of the effective activation energy}
As demonstrated above, for the strong silica, the steady-state configurations are distorted while the Si-O bond length is preserved, although elongation and compression occur in SiO$_4$ units and their network. Moreover, the role of density in the dynamics is different from that in fragile liquids \cite{FurukawaG1}. Therefore, the concept of the effective density or particle size (in the sense discussed for fragile liquids) is no longer applicable. Let us recall that in supercooled silica the structural relaxation proceeds mainly by rotational rearrangements of the SiO$_4$ tetrahedra around immobile Si atoms \cite{Buchenau-Zhou-Nucker-Gilroy-Phillips,Saksaebgwijit-Heuer}. Such rotational rearrangements occur as less-cooperative Arrhenius-like activation events that are coupled to bond-breaking and reformation processes. At equilibrium, for example, the activation energies for the rotations of a SiO$_4$ tetrahedral unit around any three-fold axis are almost equivalent. 
However, under a flow field, this is not the case due to the shear-distortions of a network and its units (elongation and compression by $\gamma=\dot\gamma\hat\tau_\alpha$): For rotations around the elongation axis, the resistance forces and the rotational displacements necessary for a transition to a different configuration, which are both perpendicular to the rotational axis, are reduced by approximately $\gamma$. Thus, for this event, the activation energy is decreased by a similar degree; two ideal cases are schematically shown in Fig. 6. In contrast, for rotations around the compression axis, the opposite occurs, that is, the activation energy is increased by approximately $\gamma$. Consequently, for a distorted SiO$_4$ network, the ``energy-landscape'' is also distorted to a  similar degree: rotational events with lower activation energies occur more easily than at equilibrium, thereby determining the structural relaxation. Note that the SiO$_4$ network structures in practice are rather heterogeneously distorted under significant thermal fluctuations, but the present qualitative argument should still be valid in the average sense. 

\begin{figure}[bt] 
\includegraphics[width=.45\textwidth]{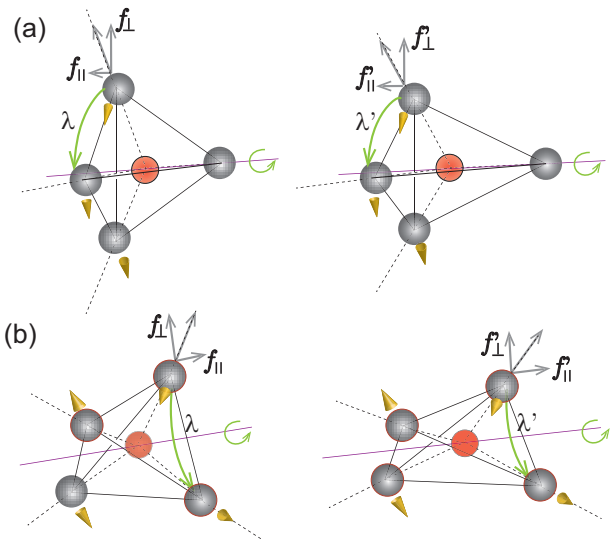}
\caption{ 
(Color online) A schematic of the rotation of a SiO$_4$ tetrahedral unit. (a) (Left) Rotation around a three-fold axis at equilibrium. The  activation energy is approximately given by $\Delta E_0 \sim  |{\mbox{\boldmath$f$}}_\bot|\lambda$. Here, ${\mbox{\boldmath$f$}}_{||}$ and  ${\mbox{\boldmath$f$}}_{\bot}$ refer to the covalent forces parallel and perpendicular to the rotational axis, respectively, and $\lambda$ is the rotational  displacement. (Right) Rotation of a distorted unit with smaller resistance force ${\mbox{\boldmath$f$}'}_\bot$ and the rotational displacement $\lambda'$ (and thus with a smaller activation energy) than those of the left unit.  (b) (Left) Rotation around a two-fold axis at equilibrium. (Right) Rotation of a distorted unit with a smaller activation energy than that of the left unit.   In (a) and (b), the displacement directions of O atoms are represented by yellow cones.
}
\label{Fig}
\end{figure}

In this qualitative perspective, the effective activation energy can be assumed to be given by  
\begin{eqnarray}
\Delta E_{\rm eff}  = \Delta E_0(1-b \gamma), 
\end{eqnarray}
where $\Delta E_0$ is the activation energy in equilibrium and $b$ is a constant of the order of unity. Similar to Eq. (\ref{mapping_fragile}), the relaxation time under the shear flow, $\hat\tau_\alpha$, is asssumed to be mapped onto the equilibrium one, $\tau_\alpha$, according to  
\begin{eqnarray}
{\hat \tau}_\alpha(T,\dot\gamma;\Delta E_0) = \tau_\alpha (T;\Delta E_{\rm eff}), \label{mapping_strong}
\end{eqnarray}
where the flow effect is incorporated by $\Delta E_{\rm eff}$. Assuming the usual Arrhenius form of $\tau_\alpha$, we obtain 
\begin{eqnarray}
{\hat \tau}_\alpha(T,\dot\gamma;\Delta E_0) = {\tau}_0 \exp \biggl[ \dfrac{\Delta E_0(1-b\gamma)}{T} \biggr], \label{mapping_strong}
\end{eqnarray}
where $\tau_0$ is a microscopic time scale. 
For $\gamma\ll1$, by expanding Eq.(\ref{mapping_strong}) in $\gamma$, we obtain 
\begin{eqnarray}
\hat\tau_\alpha(T, \dot\gamma;\Delta E_0) \cong \dfrac{\tau_\alpha(T;\Delta E_0)}{1+ b\dot\gamma \tau_\alpha \dfrac{\Delta E_0}{T}}
\end{eqnarray} 
Thus, the crossover shear rate is given by  
\begin{eqnarray}
\dot\gamma_{\rm c} \cong \biggl(\dfrac{\Delta E_0}{T}\tau_\alpha \biggr)^{-1},  \label{crossover_strong}
\end{eqnarray}
which is significantly smaller than $1/\tau_\alpha$ at lower temperatures. 
As shown in Figs. 5(c) and (d), this simple prediction is highly consistent with the simulation results of the BKS and CP models. 
In experimental studies of silicate melts with $\Delta E_0\sim 10^5$K at $T\sim10^3$K \cite{Webb-Dingwell}, the onset of non-Newtonian rheology is observed at $\dot\gamma \tau_\alpha \sim 10^{-2}$, which is consistent with our prediction, Eq. (\ref{crossover_strong}).

 \section{Concluding remarks}
 
In this study, we have investigated a difference in rheology between fragile and network-forming strong liquids using molecular dynamics simulations for four different glass formers. 
Based on our findings, we have proposed a simple shear thinning model for network-forming strong liquids, which is different from the previously proposed model for fragile liquids \cite{FurukawaS3}, and checked the validity of the models by assessing the crossover shear rate $\dot\gamma_{\rm c}$. In a high-density fragile liquid, the system properties are controlled by the repulsive interactions at nearest-neighbor distances and the density plays an essential role in blocking the rearrangement motions \cite{FurukawaG1,FurukawaG2}. Applying the average shear stress changes the equilibrium situation. That is, the particles can overlap more than at equilibrium along the compression axis, slightly reducing the effective core size or density and leading to a significant decrease of the relaxation time. In contrast, in a network-forming strong liquid, the role of density in the dynamics and in rhelogy is different from that in fragile liquids. Furthermore, shear distortions occur very differently from fragile liquids: in a silica, which is the prototype of strong liquids, shear distortions in accordance with the flow symmetry occur in the SiO$_4$ units. This structural distortion should lead to a distortion of the energy landscape to a  similar degree: for $\dot\gamma\gtrsim \dot\gamma_{\rm c}$, rotational rearrangements with lower activation energies are significantly enhanced, resulting in an acceleration of the structural relaxation.  We stress that the structural configurations undergo shear distortions in accordance with the flow symmetry, but this is not the case for the forms of the pair interactions. We have argued here that this mismatch effectively reduces the density and the activation energy for fragile and network-forming strong liquids, respectively.   
  
Finally, we note the following points: (I) Eqs. (\ref{mapping_fragile}) and (\ref{mapping_strong}) can be approximately valid even in strongly non-linear (shear-thinning) regime. Although homogeneous states have been assumed in the argument developed here, nonlocal effects such as anisotropic shear banding are pronounced in strongly non-linear regime. Such effects should be incorporated in the rheology model. 
(II) In this study, the considered systems are supercooled liquids, where thermal fluctuations exert important effects. However, these effects are irrelevant in amorphous states. In amorphous or glass states, the essential links between the shear distortion of the microscopic configurations and the nonlinear rheological properties have been intensively studied in Refs. \cite{Zaccone1,Zaccone2}. 
At this stage, it is unclear how our approach for liquids states can be related to the amorphous rheology. We will examine points (I) and (II) in future work.

\thanks

This work was supported by KAKENHI (Grant No. 26103507, No. 25000002, and No. 20508139) and the JSPS Core-to-Core Program ``International research network for non-equilibrium dynamics of soft matter''. 

\appendix

\section{Simulation Models}

In this study, we used four simple and popular model glass-forming binary mixtures: two models for strong glass formers, namely, the van Beest-Kramer-van Santen (BKS) \cite{van Beest-Kramer-van Santen} model and the Coslovich-Pastore (CP) \cite{Coslovich-Pastore} model, and two models for fragile glass formers, namely, the Kob-Andersen (KA) \cite{Kob-Andersen} and the Bernu-Hiwatari-Hansen (BHH) soft-sphere \cite{Bernu-Hiwatari-Hansen} models. These models were all simulated using Lee-Edwards periodic boundary conditions with a Gaussian thermostat \cite{RapaportB}. Here, we describe the details of these model systems.

{\it The strong BKS model.---} 
The BKS model has been extensively studied to investigate the structural and dynamical properties of amorphous and supercooled silica (Si{$\rm O_2$}) \cite{Vollmayr-Kob-Binder,Vogel-Glotzer,Horbach-Kob1,Horbach-Kob2,Horbach-Kob-Binder,Saksaebgwijit-Heuer,Saika-Voivod_Poole_Sciortino}, which is the prototypical strong glass-former. 

The interaction potential of the BKS model is given by 
\begin{eqnarray}
U_{\mu\nu}^{\rm BKS}(r)= \dfrac{q_\mu q_\nu e^2}{r} +A_{\mu\nu}\exp(-B_{\mu\nu}r) - \dfrac{C_{\mu\nu}}{r^6},  
\end{eqnarray}
where $r$ is the distance between two ions and $\mu,\nu= $Si, O. 
According to Refs. \cite{van Beest-Kramer-van Santen,Vollmayr-Kob-Binder}, the parameters are as follows: $A_{\rm Si Si}$=0 (eV), $A_{\rm Si O}$=18003.7572 (eV), $A_{\rm OO}$=1388.7730 (eV), $B_{\rm Si Si}$=0(\AA$^{-1}$), $B_{\rm Si O}$=4.87318(\AA$^{-1}$), $B_{\rm O O}$=2.76000(\AA$^{-1}$),  $C_{\rm Si Si}$=0.0 (eV\AA$^{-6}$), $C_{\rm Si O}$=133.5381 (eV\AA$^{-6}$), and $C_{\rm OO}$=175.0000 (eV\AA$^{-6}$). The partial charges are $q_{\rm Si}=2.4$ and  $q_{\rm Si}=-1.2$, and $e^2$ is given by 1602.19/4$\pi$8.8542 (eV\AA). 
The Coulombic part is usually evaluated via the Ewald summation technique, which is time-consuming. Here, to reduce the computational cost, instead of using the original BKS model, we used its simplified version, in which the Coulombic interaction is approximated by the finite-range potential given by \cite{Wolf,Carre-Berthier-Horbach-Ispas-Kob} 
\begin{eqnarray}
\dfrac{q_\mu q_\nu e^2}{r} \rightarrow {q_\mu q_\nu e^2}\biggl[\biggl(\dfrac{1}{r}-\dfrac{1}{r_{\rm c}}\biggr) +\dfrac{1}{r_{\rm c}^2}(r-r_{\rm c}) \biggr]. 
\end{eqnarray}
The potential is truncated at $r=r_{\rm c}$ while satisfying charge neutrality. In Ref. \cite{Carre-Berthier-Horbach-Ispas-Kob}, it was shown that with an appropriate choice of $r_{\rm c}$, this treatment leads to a close quantitative agreement between the truncated and non-truncated Coulombic interactions. In the present study, following Ref. \cite{Carre-Berthier-Horbach-Ispas-Kob}, we set $r_c=10.17$ (\AA). The masses of the Si and O ions were $m_{\rm Si}=$4.6638$\times 10^{-23}$ (g) and $m_{\rm O}=$2.6568$\times 10^{-23}$ (g), respectively. We fixed the mass density at 2.37 (g/cm$^3$). 
The unit length and time were $r_0=2.84$ (\AA) and $t_0=1.98\times 10^{-13}$ (s), respectively. The temperature was measured in units of 0.601 (eV)/$k_{\rm B}$=6973.9 (K), where $k_{\rm B}$ is the Boltzmann constant. 
For the main analysis, the total number of ions is $N=N_{\rm Si}+N_{\rm O}=$9000 with $N_{\rm O}/N_{\rm Si}=2$. Thus, the particle number density is $N/V=1.632$, and the linear dimension of the system is $L=17.67$ (corresponding to 50.18\AA).

\begin{figure}[bt] 
\includegraphics[width=.4\textwidth]{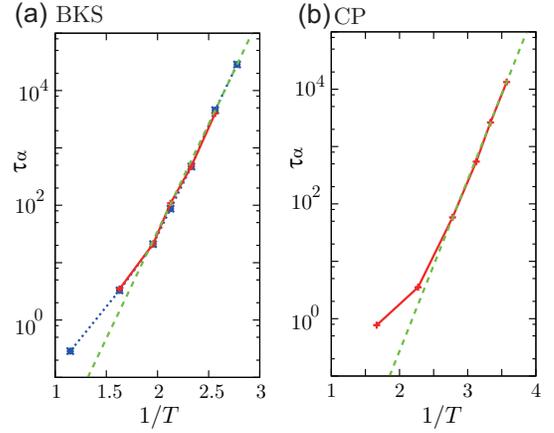}
\caption{ 
(Color online) Temperature dependence of $\tau_\alpha$ for the BKS (a) and CP (b) models (red solid line). 
In (a), the data obtained in Ref. \cite{FurukawaG1} are shown by the blue dotted line. At a lower temperature, $\tau_\alpha$ can be fitted to the Arrhenius form $\tau_0 \exp(\Delta E_0/T)$ (green dashed line), where $\tau_0=1.18\times 10^{-6}$ and , and $\Delta E_0=8.61$ and 6.87 for the BKS and CP models, respectively. }
\label{FigA1}
\end{figure}

{\it The strong CP model.---} The Coslovich-Pastore (CP) model \cite{Coslovich-Pastore} is a binary mixture of two species of particles: $A$ and $B$. The interaction potential is given by   
\begin{eqnarray}
U_{\mu\nu}^{\rm CP}(r)= \epsilon_{\mu\nu}\biggl[ \biggl(\dfrac{\lambda_{\mu\nu}}{r}\biggr)^{12}-(1-\delta_{\mu\nu})\biggl(\dfrac{\lambda_{\mu\nu}}{r}\biggr)^{6}\biggr],  
\end{eqnarray} 
where $\mu,\nu= A,B$, $\epsilon_{AB}=24\epsilon_{AA}$, $\epsilon_{BB}=\epsilon_{AA}$, $\lambda_{AB}=0.49 \lambda_{AA}$, $\lambda_{BB}=0.85 \lambda_{AA}$ and $r$ is the distance between two particles. 
Here, $\delta_{\mu\nu}$ is the Kronecher delta. 
According to Ref. \cite{Kim-Saito}, the potential is truncated at $r=2.5\lambda_{\mu\nu}$. 
The temperature $T$ was measured in units of $\epsilon_{AA}/k_B$.  
We held the particle number density constant at $N/V =1.655/\lambda_{A}^{3}$, where $N=N_A+N_B=9000$ with $N_B/N_A=2$, and $V$ is the system volume. The space and time units were $\lambda_{AA}$ and $(m \lambda_{AA}^{2}/\epsilon_{AA})^{1/2}$, respectively. Then, the linear dimension of the system was $L=17.58$. In Ref. \cite{Coslovich-Pastore}, it is shown that the CP model with these parameters accurately reproduces the results of the BKS model at $\rho=2.37$ (g/cm$^3$), although  their potential forms are very different. Thus, in the main text, particles A and B are referred to as Si and O atoms, respectively.  In Fig. \ref{FigA1}, we plot the temperature dependence of $\tau_\alpha$ for the BKS and CP models. 

\begin{figure}[hbt] 
\includegraphics[width=.4\textwidth]{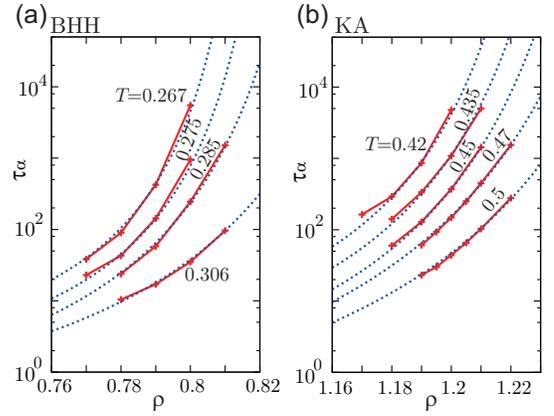}
\caption{ 
(Color online) Density dependence of $\tau_\alpha$ at equilibrium for various temperatures for the BHH (a) and KA (b) models. The data can be fitted to the form $\tau_1 \exp\{\Gamma [\rho/(\rho_{\rm c}-\rho)]^{\zeta}\}$. In this study,  we set $\zeta=1$, which gives the Vogel-Fulcher-Tamman form. However, for the present narrow density range, other values (for example, $\zeta=2$) can also fit the data.   }
\label{FigA2}
\end{figure}

{\it The fragile KA model.---} The Kob-Andersen (KA) model \cite{Kob-Andersen} is a binary mixture composed of large ($A$) and small ($B$) particles of equal masses, $m_A=m_B=m$. The interaction potential is given by   
\begin{eqnarray}
U_{\mu\nu}^{\rm KA}(r)=4\epsilon_{\mu\nu}\biggl[ \biggl(\dfrac{\lambda_{\mu\nu}}{r}\biggr)^{12}-\biggl(\dfrac{\lambda_{\mu\nu}}{r}\biggr)^{6}\biggr] - U_{\mu\nu}^0,  
\end{eqnarray}
where $\mu,\nu= A,B$, $\epsilon_{AB}=1.5\epsilon_{AA}$, $\epsilon_{BB}=0.5\epsilon_{AA}$, $\lambda_{AB}=0.8 \lambda_{AA}$, $\lambda_{BB}=0.88 \lambda_{AA}$ and $r$ is the distance between two particles. The potential is truncated at $r=2.5\lambda_{\mu\nu}$ and $U_{\mu\nu}^0$ is chosen to satisfy  $U_{\mu\nu}^{\rm KA}(2.5\lambda_{\mu\nu})=0$. The temperature $T$ was measured in units of $\epsilon_{AA}/k_B$.  
We held the particle number density constant at $N/V =1.2/\lambda_{A}^{3}$, where $N=N_A+N_B=9000$ with $N_B/N_A=2$, and $V$ is the system volume. The space and time units were $\lambda_{AA}$ and $(m \lambda_{AA}^{2}/\epsilon_{AA})^{1/2}$, respectively. Then, the linear dimension of the system was $L=31.07$. 

{\it The fragile BHH model.---} 
The Bernu-Hiwatari-Hansen model \cite{Bernu-Hiwatari-Hansen} is a binary mixture of large ($A$) and small ($B$) particles interacting via the soft-core potentials given by 
\begin{eqnarray}
U_{\mu\nu}^{\rm BHH}(r)=\epsilon \biggl(\dfrac{\lambda_{\mu\nu}}{r}\biggr)^{12}, 
\end{eqnarray}
where $\mu,\nu= A,B$, $\lambda_{\mu\nu}=(\lambda_{\mu}+\lambda_{\nu})/2$, $\lambda_{\mu}$ is the particle size, and $r$ is the distance between two particles. The mass and size ratios are $m_{B}/m_{A}=2$ and $\lambda_{B}/\lambda_{A}=1.2$, respectively. The units for the length and time are $\lambda_A$ and $({m_{A}\lambda_{A}^{2}/\epsilon})^{1/2}$, respectively. The total number of particles was  $N=N_{A}+N_{B}=8000$ and $N_A/N_B=1$.
The temperature $T$ was measured in units of $\epsilon/k_{\rm B}$. The fixed particle number density and the linear dimension of the system were $N/V =0.8/\lambda_{A}^{3}$ and $L=36.84$, respectively. 
In Fig. \ref{FigA2}, we plot the density dependence of $\tau_\alpha$ for the BHH and KA models.

\pagebreak


\begin{thebibliography}{70}
\bibitem{comment1}
In usual fragile (low molecular weight) glass formers, the packing density is significantly high: The particles are close to each other, and the steric repulsive interactions, which properly or effectively obey the additive forms, at the nearest neighbor distances dominate. In such cases, the concept of the density or the particle size can be sufficiently established (in the sense of item (ii)). Additionaly, as for item (iii), the density dependence of $\tau_\alpha$ of fragile glass formers are much steeper than those of strong glass formers.
\bibitem{comment_relaxation}
$\tau_\alpha$ is defined as the relaxation time of the shear stress autocorrelation function: $H(t)=\langle {\hat\sigma}_{xy}(t){\hat\sigma}_{xy}(0)\rangle/L^{d}T$, where $d$ is the spatial dimensionality, $L$ is the system size, and ${\hat\sigma}_{xy}$ is the $xy$-component of the shear stress tensor \cite{Hansen-McdonaldB}. $\tau_\alpha$ is determined by fitting the long-term behavior of $H(t)$ to the Kohlrausch-Williams-Watts form: $G_0 \exp[-(t/\tau_\alpha)^\psi]$, where $G_0$ is the plateau modulus and $\psi$ is the exponent of nonexponential decay.
\bibitem{comment2}
We mean the dynamics where the effects of the affine part of the particle displacements are subtracted from the total dynamics.
\end{thebibliography}
\end{document}